\documentclass{elsart3p}

\usepackage{amssymb,graphicx}

\begin{document}

\begin{frontmatter}



\title{Asymptotic expansions for the escape rate of stochastically perturbed unimodal maps}


\author{Carl P. Dettmann} and
%\address{Department of Mathematics, University of Bristol, University Walk, Bristol, BS8 1TW, UK}
\author{Teil B. Howard}
\address{Department of Mathematics, University of Bristol, University Walk, Bristol, BS8 1TW, UK}

\begin{abstract}
The escape rate of a stochastic dynamical system can be found as an expansion in powers of the noise strength. In previous work the coefficients of such an expansion for a one-dimensional map were fitted to a general form containing a few parameters. These parameters were found to be related to the fractal structure of the repeller of the system. The parameter $\alpha$, the ``noise dimension'', remains to be interpreted. This report presents new data for $\alpha$ showing that the relation to the dimensions is more complicated than predicted in earlier work and oscillates as a function of the map parameter, in contrast to other dimension-like quantities.

\end{abstract}

\end{frontmatter}

% main text
\section{Introduction}
%\label{}
Physical systems all experience noise to some degree. Whether from unknown degrees of freedom, thermal fluctuations, quantum fluctuations or many other causes the dynamics are affected in all systems from neuronal systems~\cite{10,11} to cosmology~\cite{12}. Differential equations governing physical systems are idealised and in reality will be affected at some scale. This motivates the study of systems which evolve with noise, specifically maps with additive noise.

The problem of whether a particle escapes or at what rate particles escape from an open system is another subject that has many physical applications, from an asteroid escaping a planet~\cite{13} in celestial mechanics to the chemical reactions that occur in transition state theory~\cite{14}. The study of billiards has expanded to include open systems~\cite{17,19} and open optical billiards may have industrial applications~\cite{19a}. All of these physical systems may also experience noise. 

This paper considers asymptotic expansions for the escape rate from one dimensional unimodal maps with small additive noise. One-dimensional maps have shown to be models of properties such as universality~\cite{24,25}, intermittency~\cite{26,27} and deterministic diffusion~\cite{22,23}. The effect of noise on the properties of one-dimensional maps has been well studied~\cite{29,31,33,n2,37,n1} particularly in the vicinity of bifurcation and crisis points~\cite{35,34,36}. Properties such as escape rates and lifetimes of transient behaviours have been shown to have a nonmonotonic dependence on noise strength~\cite{32}. In contrast to most of the previous work we discuss the high order terms in the weak noise expansion which determine the rate of convergence or divergence and hence the usefulness of the expansion.

The escape rate from a one-dimensional map with additive noise can be calculated as an expansion in terms of the noise strength using information about the periodic orbits of the corresponding deterministic map. Different approaches exist~\cite{a,b} but the most accurate results are obtained using a numerical method described in~\cite{1}. This method accurately yields a large number of terms in the expansion but takes a large computational time. In~\cite{2} the expansion is reformulated into a new asymptotic form where the parameters are related to properties of the system including the complex fractal dimension. Many physical systems show complex dimensions. Ref.~\cite{16} discusses the structure of the bronchial tree in the mammalian lung and Ref.~\cite{16a} reviews a number of other applications including the rate of escape from a stable attractor surrounded by an unstable limit cycle as a function of noise strength and the formation of black holes with mass smaller than the Chandrasekhar limit.

However, the parameter $\alpha$, the ``noise dimension'', remains to be found. Finding the significance of this parameter could enable the expansion of the escape rate to be calculated more easily or to higher order. Conversely, it could also allow the properties of a system to be inferred from experimental data for the escape rate.

In~\cite{2} it is suggested that the noise dimension may be related to the generalised Renyi dimensions of the fractal repeller which are known to change as the map changes. This paper uses numerical techniques to calculate the noise dimension for a range of maps and compares the data to the Renyi dimensions in an attempt to find the significance of the remaining parameter. Generalisations to higher dimension or continuous time systems are likely, however, this will require a greater understanding of one-dimensional systems.

Section 2 reviews some results from periodic orbit theory which lead us to the escape rate for deterministic and stochastically perturbed systems. In Section 3 the theory of asymptotic expansions is generalised to include complex ``exponents'' and the new formulation of the expansion of the escape rate is introduced. Numerical results for the noise dimension are found and discussed in Section 4 and finally our conclusions are given in Section 5.

\section{Periodic orbit expansions for the escape rate}

The systems considered include unstable and chaotic trajectories, particularly when there is noise, so we study the evolution of probability densities rather than individual trajectories. The evolution of a density of particles moving in a deterministic dynamical system is governed by the Perron-Frobenius evolution operator~\cite{15}. If the dynamics are described by a discrete time map $x_{n+1}=f(x_n)$ where $x_n$ is a real number, then the probability density, $\phi(x)$, evolves as
\begin{equation}\label{densev}
\phi_{n+1}(y)=({\cal L}\circ\phi_n)(y)=\int\phi_n(x){\cal L}(y,x)dx
\end{equation}
with the evolution operator ${\cal L}$ given by
\begin{equation}
	{\cal L}(y,x)=\delta(y-f(x)).
\end{equation}
The Perron-Frobenius evolution operator can also be generalised to higher-dimensional and continuous systems.

For a sufficiently uniformly chaotic system with a uniform distribution of initial conditions the number remaining in the considered region at long time $n$ is proportional to $e^{-\gamma n}$, where $\gamma$ is the escape rate. The escape rate of an open deterministic system can easily be calculated from the leading eigenvalue, $\nu$, of the evolution operator using the relation, $\gamma=-\ln\nu$. Other properties such as averages, Lyapunov exponents and correlation functions are also related to the leading eigenvalue of the evolution operator. This eigenvalue is simply the leading root of the spectral determinant of the linear operator which can be expressed as an expansion in terms of powers of traces of the evolution operator as shown below.
\begin{eqnarray}
& 0& =\det(1-{z\cal L})=\exp{\rm tr}\ln(1-z{\cal L})\nonumber \\
& &=1-z{\rm tr}{\cal L}-\frac{z^2}{2}({\rm tr}{\cal L}^2-({\rm tr}{\cal L})^2)+\cdots .
\end{eqnarray}
Here $z$ is the inverse eigenvalue and 
\begin{equation}
{\rm tr}{\cal L}^n=\int dx{\cal L}^n(x,x)=\int dx\delta(x-f^n(x)).
\end{equation}
The only values of $x$ which contribute to this integral are those that lie on periodic orbits. Periodic orbit theory provides a simple formula to calculate these traces using information about the periodic points of the system~\cite{7,9}.

In the case of a stochastic dynamical system where the system experiences some additive noise the equivalent evolution operator is the Fokker-Planck evolution operator \cite{7}. If the dynamics are described by a discrete time map $x_{n+1}=f(x_n)+\sigma \xi_n$ where $\sigma$ is the noise strength and $\xi_n$ are instances of a random variable with normalized distribution $p(\xi)$ then the probability density, $\phi(x)$, evolves as Eq. (\ref{densev}) but now the evolution operator ${\cal L}$ is given by
\begin{equation}
{\cal L}(y,x)=\delta_\sigma(y-f(x)) 
\end{equation}
where
\begin{equation}
\delta_\sigma(x)=\int\delta(x-\sigma\xi)p(\xi)d\xi=\frac{1}{\sigma}p\left(\frac{x}{\sigma}\right).
\end{equation}
As in the deterministic case, generalisations to higher-dimensional and continuous time systems are possible. Equivalent results are shown in \cite{28} where the trace formula is found for a noisy flow.

For stochastic systems the problem of finding the escape rate can also be reduced to finding the traces. Here we have
\begin{eqnarray}
{\rm tr}{\cal L}^n=\int \delta_\sigma(x_1-f(x_0))...\delta_\sigma(x_0-f(x_{n-1})) \nonumber \\
\times dx_0...dx_{n-1}
\end{eqnarray}
but now there is no simple formula that can be used from periodic orbit theory. Two analytical approaches have used Feynman diagrams~\cite{a} and smooth conjugations~\cite{b} to express the traces as expansions in terms of the noise strength. However, they have only obtained results up to order $\sigma^4$. Instead a different more numerical approach may be taken which expresses the evolution operator as a matrix which is calculated in the vicinity of each periodic orbit to find the traces and then obtains the eigenvalue as an expansion in terms of the noise strength \cite{1}. When used with high precision arithmetic this method can obtain results of up to order 64 in the noise strength~\cite{2}. With such a high order expansion obtained to a high degree of accuracy it is possible to look at the asymptotic form of the late terms. Reformulating the expansion in terms of new parameters could allow it to be calculated more efficiently or to a higher order.

Previous work in~\cite{2} proposed one such new formulation. By studying one particular map most of the new parameters were found but a parameter known as $\alpha$ remained. However a tentative relation between $\alpha$ and the dimension was proposed. The aim of this paper is to present new data for $\alpha$ by looking at different maps and to show that the behaviour is more complex than predicted in the previous work. A brief description of the derivation of the parameters follows but a more thorough explanation is given in the original paper~\cite{2}.
	
\section{The expansion for the escape rate in powers of 
the noise strength}

The expansion for a smooth map with symmetric noise is fitted in~\cite{2} to the following asymptotic form,
\begin{equation}\label{exp}
\nu(\sigma)=\sum_{m}\nu_{2m}\sigma^{2m}
\end{equation}
where 
\begin{equation}\label{nu}
\nu_{2m}\sim \frac{1}{\sigma_0^{2m}}\sum_{n=-\infty}^\infty c_n (m+\alpha + in\beta)!
\end{equation}
Here $\sigma$ is the noise strength and $\nu_{2m}$ are coefficients that can be obtained using the local matrix approach~\cite{1}.

Eq. (\ref{nu}) without the complex $in\beta$ (and hence sum over $n$) is familiar from the theory of asymptotic series for integrals of the form $\int\exp (g(u)/\sigma)h(u)du$ \cite{D,8}. The quantity 
\begin{equation}
\frac{\nu_{2m}\sigma_0^{2m}}{(m+\alpha)!}
\end{equation} 
would normally be expected to tend to a constant as $m$ tends to infinity, however, in~\cite{2} it was found to exhibit oscillatory behaviour. The complex $in\beta$ were added in order to replicate this behaviour. 

The parameter $\sigma_0$, called the singulant, is given by the variation of $g(u)$ between the point defining the asymptotic expansion and the closest critical point~\cite{D}. Here this corresponds to finding the probability of returning to the repeller from the critical point of the map~\cite{2}. The integration variable $u$ is in our case the whole path $\{x_i\}$ and a maximally probable path is then a critical point in this infinite-dimensional space; see Section 4 for the explicit calculation.

The remaining parameters, $\alpha$, $\beta$ and $c_n$, were found in~\cite{2} to be related to the fractal structure of the set of repelling points of the map. We motivate this connection as follows: in \cite{6} a similar analytic expansion was found for the electrostatic potential near the edge of a middle-third Cantor set with a uniform charge on it. The middle-third Cantor set contains two copies of itself scaled down by a factor of three so has dimension $d=\ln 2/\ln 3$. The value of the potential a distance $\xi$ from the edge of the set is given by a series of $\xi^{d-1}$ where the $d$ values are the complex solutions of the equation that describes the dimension of the set, $3^d=2$.
\begin{equation}
	d=\frac{{\ln}\,2}{{\ln}\,3}+\frac{2\pi in}{{\ln}\,3}=\alpha+in\beta,
\end{equation}
where $n$ is an integer, $\alpha$ is the dimension and $\beta$ is related to the spatial scaling factor of the fractal. Oscillatory terms arise in this expansion as a result of the complex values of $d$ corresponding to $n\neq0$. In this expansion the coefficients, $c_n$, have a rapid exponential decay so only $|n|<2$ are required. Complex fractal dimensions such as these appear in many physical applications~\cite{16,16a}.

Looking back at the original expansion defined by (\ref{exp}) and (\ref{nu}) we can analogously relate the values of $\alpha$ and $\beta$ to the properties of the fractal repeller of the map. The middle-third Cantor set is an exactly self-similar fractal, however, the repeller of the map may not be exactly self-similar and may be a multifractal. In this case the spatial scaling factor would vary with position. In \cite{2} $\beta$ is shown to be related to the spatial scaling factor at the most probable point of return to the considered region from the critical point and is given by
 \begin{equation}
	\beta=\frac{2\pi}{{\ln}\,f'(x_{r})}
\end{equation}
where $f$ is the map and $x_r$ is the point of return. However, for the parameter $\alpha$, the noise dimension, it is not so simple. As the repeller may be a multifractal, we need the concept of generalised dimension. The Renyi dimension of a fractal covered by boxes of size $\epsilon$ is defined as
\begin{equation}
	D_q=\frac{1}{q-1}\lim_{\epsilon \rightarrow 0}\frac{{\ln} \,\sum_{i} p_i^q}{{\ln}\, \epsilon}
\end{equation}
where $p_i$ is the amount of the fractal in box $i$~\cite{7}. The middle-third Cantor set is exactly self-similar so has dimension $d={\ln}\,2/{\ln}\,3$ for all values of $q$. However, when the repeller for the map is not exactly self-similar there are infinitely many different dimensions. In this context it is not clear which dimension is related to $\alpha$.

\section{Results: $\alpha$ vs. Renyi dimensions}

Recall from the previous section that the escape rate of a stochastically perturbed map may be fitted to the form given by (\ref{exp}) and (\ref{nu}). Explicit expressions can be found for the parameters $\sigma_0$ and $\beta$. However, the parameters $\alpha$ and $c_n$ remain to be found. In this section we calculate $\alpha$ and $c_n$ numerically by fitting the new form to the initial expansion and minimising the least squares error after finding the values of the other parameters. Following the equation for the potential near the middle-third Cantor set we assume that the $c_n$ values decay exponentially and also use only $|n|<2$ which is found to be justified when the first $c_n$ values are calculated. We then compare the data to the dimensions in an attempt to better understand the significance of $\alpha$.

\begin{figure}[h]
\begin{center}
	\includegraphics[width=4.5cm]{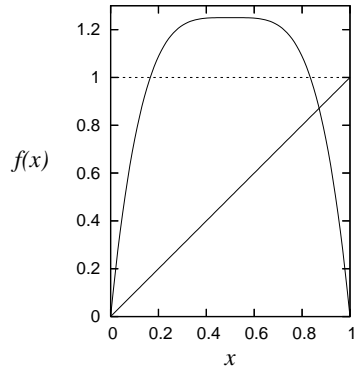}\\
	\caption{The map defined by (\ref{smap1}) when $a=0.8$.\label{map}}
\end{center}
\end{figure}

We have considered systems where the noise is additive and the noise strength is given by  $\sigma$, so an individual trajectory will evolve with time as
\begin{equation}
	x_{n+1}=f(x_n)+\sigma\xi_n,
\end{equation}
where the one-dimensional, real variable $x_n$ is the position at $t=n$, $f$ is the map and $\xi_n$ are instances of an independent and identically distributed random variable. Previous work~\cite{1,2} has concentrated on the map where
\begin{equation}\label{smap1}
	f(x)=\frac{1}{a}(1-(1-2x)^4)
\end{equation}
and $a=0.8$.
The map can be changed by varying the map defining parameter, $a$, between $0$ and $1$. Fig.~\ref{map} shows this map for $a=0.8$.
These specific maps were chosen because it is simple to find the inverse branches and they have complete binary symbolic dynamics which means periodic orbits are easy to enumerate using inverse iteration, however, the method does not require complete symbolic dynamics. With this map we are interested in the $[0,1]$ region and in order to escape a particle must leave this region. The noise chosen here is Gaussian with distribution 
\begin{equation}
	p(\xi)=\frac{e^{-\xi^{2}/2}}{\sqrt{2\pi}},
\end{equation}
however, any other peaked smooth symmetric noise distribution would be appropriate including those that are space dependent. 

For small $\sigma$ the probability of returning to the repeller from the critical point of the map where $f'(x)=0$ is given by (ignoring algebraic dependence)
\begin{equation}
\exp\left ( -\frac{1}{2} \left ( \frac{\sigma_0}{\sigma}  \right )^2\right ) 
\end{equation}
where we identify $\sigma_0$ with the singulant that appears in (\ref{nu}) \cite{2}. In this limit the probability is given to leading order by the probability of the most likely trajectory that returns to the repeller from the critical point. For the map defined by (\ref{smap1}) the critical point is $x_c=0.5$. From here the most likely point of return to the repeller is at the edge of the set. If the trajectory returns after one step $x_r=1$ or if it returns after more than one step $x_r=0$. The probability of a trajectory $\{ x_0,x_1,...,x_n\}$ is given by
\begin{equation}
\exp \left ( -\frac{\sum_{i=0}^n (x_{i+1}-f(x_i))^2}{2\sigma^2}\right ).
\end{equation}
Finding the most probable trajectory with $x_0=x_c=0.5$ and $x_n=x_r$ is equivalent to minimising the sum over all possible lengths $n$ and values of $x_1,x_2,...,x_{n-1}$. Calculating this minimum for each $n$ shows that longer trajectories are more likely. Previous work in~\cite{2} found that when $a=0.8$ in (\ref{smap1}) the probability is maximised by the infinite trajectory $\{ 0.5, 1.00244613635157,-0.00024587488150, \\ 2.460023246\times10^{-5},-2.46015082 \times 10^{-6},\\-2.4601636 \times10^{-7}, -2.460165\times10^{-8},...,0\}$ and $\sigma_0=32.31850341240166^{-1/2}$. On this trajectory the approach to the repeller is geometric with ratio $1/10=1/f'(x_r)$ which is the inverse of the spatial scaling factor.

Using the most likely point of return, $x_r=0$, we can also find $\beta$ as follows,
\begin{equation}
	\beta=\frac{2\pi}{{\ln}\,f'(x_{r})}=\frac{2\pi}{{\ln}\left( \frac{8}{a}(1-2x_r)^3 \right)}=\frac{2\pi}{{\ln}(8/a)}.
\end{equation}

In~\cite{2} one specific case of the map was considered where $a=0.8$. When the fitting described above was carried out a value of $\alpha=-1.290$ was found. The Renyi generalised dimensions~\cite{7} were also calculated for this system and $D_\infty$ was found to be 0.2957. At this point a tentative connection between $\alpha$ and $D_\infty$ was proposed, $\alpha=-1-D_\infty$. This report presents further data for $\alpha$, as such conclusions cannot be drawn from only one piece of data. 

\begin{figure}[h]
\begin{center}
\includegraphics[width=5.3cm,angle=270]{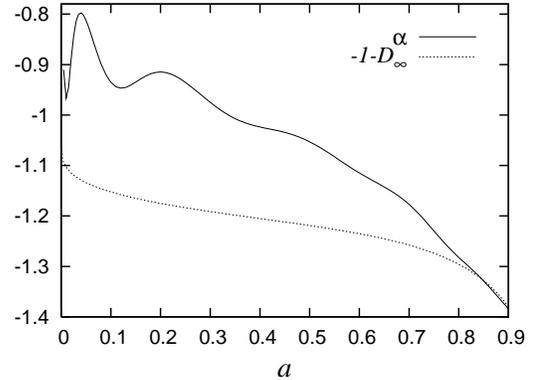}
\caption{Plot of the noise dimension, $\alpha$, and $-1-D_{\infty}$ for the map defined by (\ref{smap1}). \label{res1}}
\end{center}
\end{figure}

\begin{figure}[h]
\begin{center}
\includegraphics[width=5.3cm,angle=270]{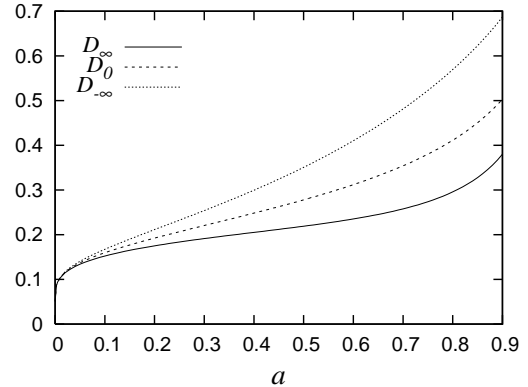}
\caption{Plot of $D_{\infty}$, $D_0$ and $D_{-\infty}$ for the map defined by (\ref{smap1}). \label{res2}}
\end{center}
\end{figure}

\begin{figure}[h]
\begin{center}
\includegraphics[width=5.3cm,angle=270]{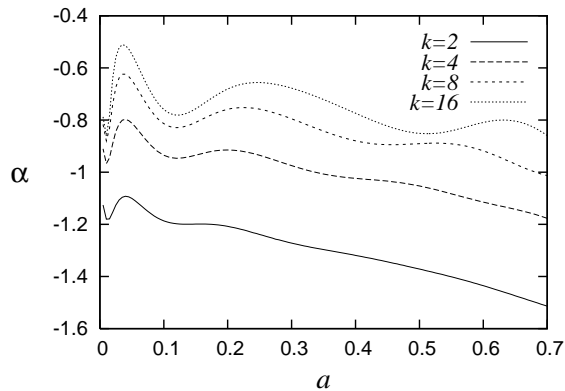}
\caption{Plot of the noise dimension, $\alpha$, for maps defined by (\ref{smap2}). \label{res3}}
\end{center}
\end{figure}

\begin{figure}[h]
\begin{center}
\includegraphics[width=5.3cm,angle=270]{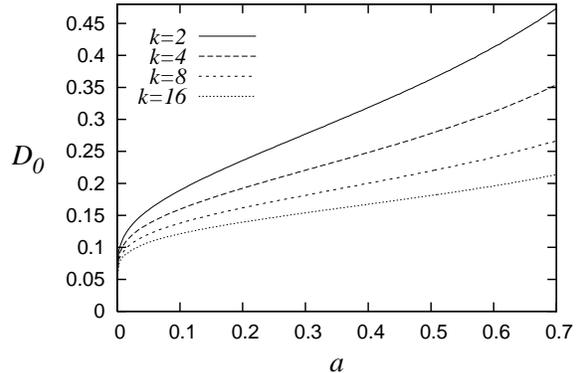}
\caption{Plot of $D_0$ for maps defined by (\ref{smap2}). \label{res4}}
\end{center}
\end{figure}

Results obtained are shown in Figs.~\ref{res1}-\ref{res4}. Fig.~\ref{res1} shows a plot of the behaviour of $\alpha$ as the map is varied along with $-1-D_{\infty}$. The data for $\alpha$ was obtained using a noise expansion with terms up to order 48 calculated using periodic orbits with length up to $12$. The new form is fitted to the initial expansion using coefficients of $\sigma^{12}$ and higher order terms to perform a least squares fit.

\nopagebreak
\nopagebreak

Comparison between $\alpha$ and $-1-D_{\infty}$ shows that the graphs are clearly different to each other and the relation proposed in~\cite{2} holds only approximately for $0.8<a<0.9$. 

As $a$ tends to zero the repeller of the map tends to that of an infinitely high tent map. The repeller of the tent map is a Cantor set, which is an exactly self-similar fractal. All the dimensions will have the same value and this value will tend to zero as the height increases. As a result of this we would expect the plot of $\alpha$ to tend to a constant as $a$ tends to zero if it were a simple function of dimension, but instead, we find it oscillates increasingly fast as it approaches zero. 

The variation of the generalised Renyi dimensions $D_{\infty}$, $D_0$ and $D_{-\infty}$ is shown in Fig.~\ref{res2}. The plot of any other Renyi dimension should also be smooth and lie between the plots of $D_{\infty}$ and $D_{-\infty}$. Examination of this plot shows that the dimensions do tend to the behaviour of the tent map as $a$ tends to zero, however, they do not exhibit behaviour as oscillatory as that of $\alpha$ and so it seems unlikely that there is a simple relationship between $\alpha$ and $D_q$ for any $q$. 

In the case of the exactly self-similar fractal all of the Renyi dimensions were identical and $\alpha$ had a simple relationship to the dimension but in the case studied here the fractal repeller is not exactly self-similar so the Renyi dimensions are all different. In this situation it is not clear which dimensions to use and this may account for the more complex behaviour.

We can also apply this method to different maps. Fig.~\ref{res3} shows data for a selection of different maps of the form
\begin{equation}\label{smap2}
f(x)=\frac{1}{a}(1-(1-2x)^k).
\end{equation}
As the order of the map increases $\alpha$ increases and the oscillations become larger. Fig.~\ref{res4} shows the corresponding variation of the dimension $D_0$ for these maps. It is clear that the variation of the dimensions with $a$ or $k$ is very different to the variation of $\alpha$. This further supports the idea that $\alpha$ is not a simple function of the dimensions. The only clear similarity is that as $k$ increases the plots get closer together, however, this is expected as the map changes less as $k$ increases.



\section{Conclusions}

The escape rate from a stochastically perturbed one-dimensional map can be given as an expansion in terms of the noise strength. Previous work~\cite{2} fitted this expansion to a new form with parameters that depend on the fractal properties of the repeller. In this work we have focused on one parameter, the noise dimension $\alpha$, which was thought to be related to the complex fractal dimension~\cite{16,16a} but remained to be found.

Although we have not found the noise dimension, $\alpha$, explicitly in terms of other known quantities we have produced new data showing how it varies as the map varies. We have calculated $\alpha$ as the map parameter is changed for the quartic map given by (\ref{smap1}) and higher order maps given by (\ref{smap2}). We have shown that any relation to the Renyi dimensions of the repeller is much more complicated than predicted. The dimensions are smooth monotonic functions of the map parameter $a$ whereas $\alpha$ oscillates.

Future work will continue to look for the significance of the noise dimension and its relation to the static and dynamic properties of the repeller. Finding this parameter could allow the expansion for the escape rate to be found more simply or allow us to draw conclusions about the properties of a system from experimental measurements of the escape rate. The data presented for $\alpha$ poses many new questions, such as, what is the limiting behaviour as the height or power of the map tends to infinity and can the oscillations in $\alpha$ be described as a function of the map defining parameter $a$? We hope that answering these questions will provide new insight into the nature of the escape from stochastic dynamical systems.

\section*{Acknowledgements}
The authors are grateful for helpful discussions with Jon Keating and Martin Sieber.  TH was supported by an EPSRC doctoral training grant.

\end{document}